\documentclass[a4paper]{jpconf}

\usepackage[english]{babel}
\usepackage{amsmath,amssymb}
\usepackage[dvips]{graphicx}
\usepackage{rotating}

\def\rpv{$R_p \hspace{-1em}/\;\:\hspace{0.2em}$}

\def\lsim{\raise0.3ex\hbox{$\;<$\kern-0.75em\raise-1.1ex\hbox{$\sim\;$}}}
\def\vb#1{\vbox to #1 pt{}}

\begin{document}

\title{Spontaneous R-parity violation and the origin of neutrino mass}

\author{A.~Vicente}

\address{AHEP Group, Institut de F\'{\i}sica Corpuscular --
  C.S.I.C. \& Universitat de Val{\`e}ncia \\
  Edificio Institutos de Paterna, Apt 22085, E--46071 Valencia, Spain}

\ead{Avelino.Vicente@ific.uv.es}

\begin{abstract}

We study the phenomenology of supersymmetric models that explain neutrino masses through the spontaneous breaking of R-parity, finding strong correlations between the decays of the lightest neutralino and the neutrino mixing angles. In addition, the existence of a Goldstone boson, usually called Majoron ($J$), completely modifies the phenomenology with respect to the standard picture, inducing large invisible branching ratios and charged lepton decays, like $\mu \to e J$, interesting signals that can be used to constrain the model.

\end{abstract}

\section{Introduction}

Nowadays it is well stablished that neutrinos have non-zero masses and mixing angles. In fact, oscillation experiments \cite{Fukuda:1998mi,Ahmad:2002jz,Eguchi:2002dm,Collaboration:2007zza,KamLAND2007} have become more precise in the last years, allowing us to measure the involved parameters with good accuracy. Moreover, global fits \cite{Schwetz:2008er} have shown the existence of two large mixing angles, $\theta_{12}$ and $\theta_{23}$, one small (perhaps zero) mixing angle, $\theta_{13}$, and two different mass scales, given by the squared mass differences $\Delta m_{12}^2$ and $\Delta m_{23}^2$.

Therefore, since the minimal models do not include neutrino masses, we must enlarge them in order to explain this data. Many proposals can be found in the literature with that purpose. The most popular ones are those based on the different variations of the so-called seesaw mechanism \cite{Minkowski:1977sc,Yanagida:1979as,Gell-Mann:1980vs,Mohapatra:1979ia,Schechter:1980gr}, in which the existence of a high energy scale can naturally explain the smallness of neutrino masses. However, it is impossible to test these models directly, and only with some additional assumptions one might be able to get indirect insights thanks to forthcoming experiments \cite{Deppisch:2002vz,Blair:2002pg,Freitas:2005et,Buckley:2006nv,Deppisch:2007xu}.

Consequently, it is worth to consider alternative models, in which neutrino masses are generated at the electroweak scale. In this case one can also find many different ideas. One possibility is to get neutrino masses through radiative corrections, like in the Zee model \cite{Zee:1980ai} or in the Babu-Zee model \cite{Babu:1988ki}, two good examples of models with a rich phenomenology at current colliders like the LHC. On the other hand, in the context of supersymmetry, the violation of R-parity (\rpv) \cite{Hall:1984id} is an interesting alternative, since it can provide connections between the decay of the lightest supersymmetric particle, the LSP, and neutrino physics \cite{Porod:2000hv,Restrepo:2001me,Hirsch:2002ys,Bartl:2003uq,Hirsch:2003fe}.

There are many variations of \rpv models. The simplest versions break R-parity including explicit terms in the superpotential that break either lepton or baryon number. Following that idea, of particular interest for neutrino physics are bilinear R-parity violating models (BRpV) \cite{Hirsch:2004he}, in which lepton number violating terms $\epsilon \widehat H \widehat L$ are introduced. The mixing between the neutrinos and the rest of neutral fermions induce small neutrino masses if the dimensionful $\epsilon$ parameters are small compared to the SUSY scale, giving a TeV scale version of the seesaw mechanism.

However, there is no explanation for the smallness of the $\epsilon$ parameters, since they are SUSY conserving terms, and therefore their natural scale is the Planck scale. Spontaneous violation of R-parity (s-\rpv) \cite{Aulakh:1982yn,Masiero:1990uj} can give a natural solution, generating the bilinear terms at the electroweak scale\footnote{Recently, other models in which the bilinear terms are generated spontaneously have appeared in the literature. As a simple but interesting example, the so-called $\mu \nu$SSM \cite{LopezFogliani:2005yw} has many things in common with s-\rpv, promising a rich phenomenology at colliders.} while keeping the nice features of the simplest models. Moreover, in addition to the usual signatures related to LSP decays, the spontaneous breaking of lepton number implies the existence of a Goldstone boson, the Majoron ($J$), which can strongly modify the phenomenology. For example, as it was already shown in \cite{Hirsch:2006di,Hirsch:2008ur}, the invisible decay channel $\chi^0 \to J \nu$ can be dominant, leading to a possible confusion with the usual conserved R-parity models.

In this paper we will focus on the phenomenology associated with the decays of the lightest neutral fermion, and how we can extract information connected to neutrino physics, for different scenarios: (i) a bino-like LSP, very common in the literature for mSUGRA models, and (ii) a singlino-like LSP. In particular we will confirm the result in \cite{Hirsch:2006di,Hirsch:2008ur}, where the difficulties to distinguish the model from R-parity conserved in special regions of parameter space are shown. In connection with that, exotic muon decays like $\mu \to e J$ and $\mu \to e J \gamma$ are enhanced in the same regions of parameter space where the invisible decays of the LSP are dominant, thus providing information complementary to accelerator searches. 

The paper is organized as follows: in Section \ref{sec:model} we present the model with special emphasis on neutrino physics. In Section \ref{sec:pheno} we discuss production and decays of the lightest neutralino at the LHC. In particular, we will stress the importance of the different correlations between ratios of neutralino decay branching ratios and neutrino mixing angles. In Section \ref{sec:exotic} we study exotic muon deays involving Majorons and how to use them to constrain the model. Finally, we draw our conclusions in Section \ref{sec:con}.

\section{The model}
\label{sec:model}

\subsection{Model basics}

The model we consider \cite{Masiero:1990uj} contains three additional singlet superfields, namely, $\widehat\nu^c$, $\widehat S$ and $\widehat\Phi$, with lepton number assignments of $L=-1,1,0$ respectively. Then, taking into account two basic guidelines: (i) R-parity is conserved at the level of the superpotential, and (ii) We do not include terms with dimensions of mass, we can write down the following superpotential:

\begin{eqnarray} %
{\cal W} &=& h_U^{ij}\widehat Q_i \widehat U_j\widehat H_u
          +  h_D^{ij}\widehat Q_i\widehat D_j\widehat H_d
          +  h_E^{ij}\widehat L_i\widehat E_j\widehat H_d \nonumber
\\
        & + & h_{\nu}^{i}\widehat L_i\widehat \nu^c\widehat H_u
          - h_0 \widehat H_d \widehat H_u \widehat\Phi
          + h \widehat\Phi \widehat\nu^c\widehat S +
          \frac{\lambda}{3!} \widehat\Phi^3
\label{eq:Wsuppot}
\end{eqnarray}

It must be noticed that the introduction of the $\widehat\Phi$ superfield addresses the $\mu$-problem of supersymmetry a l\'a NMSSM \cite{Barbieri:1982eh}. Similarly, the right-handed neutrino superfield $\widehat \nu^c$ will play a similar role in the spontaneous generation of bilinear \rpv terms, through its Yukawa coupling with the left-handed neutrino superfield. For both the inclusion of the second guideline is needed. Finally, the soft supersymmetry breaking terms of this model can be found in \cite{Hirsch:2004rw}.

When electroweak symmetry breaking occurs, several scalar fields acquire vevs \cite{Romao:1992vu}. Besides the usual Higgs bosons

\begin{equation}
\langle H_d^0 \rangle = \frac{v_d}{\sqrt{2}} \quad \langle H_u^0 \rangle = \frac{v_u}{\sqrt{2}}
\end{equation}

we also have

\begin{equation}
\langle \Phi \rangle = \frac{v_{\Phi}}{\sqrt{2}} \quad \langle {\tilde \nu}^c \rangle = \frac{v_R}{\sqrt{2}} \quad \langle {\tilde S} \rangle = \frac{v_S}{\sqrt{2}} \quad \langle {\tilde \nu}_i \rangle = \frac{v_{L_i}}{\sqrt{2}}
\end{equation}

As a consequence of this vacuum structure, an effective $\mu$ term appears, $\mu = h_0 \frac{v_{\Phi}}{\sqrt{2}}$, which is naturally at the electroweak scale. Moreover, R-parity is spontaneously broken by the vevs $v_R$, $v_S$ and $v_{L_i}$, and bilinear \rpv terms $\epsilon_i = h_\nu^i \frac{v_R}{\sqrt{2}}$ are generated as well. In this model their smallness is naturally explained by the smallness of the Yukawa couplings of the neutrinos, $h_\nu^i$, which are also connected to the vevs of the left-handed sneutrinos, $v_{L_i}$, through the tadpole equations \cite{Masiero:1990uj}. Therefore, we will consider in the following that both $\epsilon_i$ and $v_{L_i}$ are small parameters compared to the usual SUSY scale.

This spontaneous breaking of lepton number leads to the appearance of a Goldstone boson, usually called Majoron, $J$. The existence of a massless Majoron is usually seen as a bad feature of models, since the first s-\rpv models \cite{Aulakh:1982yn} were ruled out by LEP and astrophysical data \cite{Raffelt:1996wa,Yao:2006px} because of the doublet nature of this Goldstone boson. However, in this model R-parity is broken by the vevs of singlets, and therefore the Majoron has singlet nature, avoiding the strong constraints. In fact, it is possible to find a simple approximation for the Majoron

\begin{equation}\label{smplstmaj}
J \simeq \big(0,0,\frac{v_{L k}}{V},0,\frac{v_S}{V},-\frac{v_R}{V}\big)
\end{equation}

where $V=\sqrt{v_R^2+v_S^2}$ and terms of order $\frac{v_L^2}{V v}$ have been neglected. As we will see, the presence of this massless particle can strongly modify the phenomenology at colliders from the standard picture, introducing invisible decay channels which can have large branching ratios. Moreover, it can also have an impact on charged lepton decays, opening the possibility to processes like $l_i \to l_j J$ and $l_i \to l_j J \gamma$.

\subsection{Neutrino masses}

Once lepton number is broken neutrinos get masses through neutralino-neutrino mixing. In the basis $(\psi^0)^T = (\tilde B^0,\tilde W^0_3,{\tilde H_d},{\tilde H_u},\nu_e,\nu_{\mu},\nu_{\tau},\nu^c,S,\tilde{\Phi})$ the $10 \times 10$ neutral fermion mass matrix can be written as \cite{Hirsch:2004rw,Hirsch:2005wd}

\begin{equation}
\mathbf{M_N} = \left(
\begin{array}{cc}
\mathbf{M_H} & \mathbf{m}_{3\times 7} \cr \vb{20} \mathbf{m}_{3\times 7}^T & 0 \cr
\end{array}
\right)
\end{equation}

The sub-block $\mathbf{M_H}$ is the mass matrix of the seven heavy states while $\mathbf{m}_{3\times 7}$ is the matrix that mixes them with the light neutrinos. Their exact expressions can be found in \cite{Hirsch:2008ur}.

Since neutrino masses are much smaller than all other fermion mass terms, one can find the effective neutrino mass matrix in seesaw approximation \cite{Hirsch:2004rw,Hirsch:2005wd}:

\begin{equation}
\boldsymbol{m_{\nu\nu}^{\rm eff}} = - \mathbf{m}_{3\times 7} \cdot \mathbf{M_H}^{-1} \cdot \mathbf{m}_{3\times 7}^T
\end{equation}

and, after some straightforward algebra, $\boldsymbol{m_{\nu\nu}^{\rm eff}}$ can be cast into a very simple form

\begin{equation}\label{meff}
-(\boldsymbol{m_{\nu\nu}^{\rm eff}})_{ij} = a \Lambda_i \Lambda_j +
     b (\epsilon_i \Lambda_j + \epsilon_j \Lambda_i) +
     c \epsilon_i \epsilon_j
\end{equation}

where $\Lambda_i = \epsilon_i v_d + \mu v_{L_i}$ are the so-called alignment parameters, that also appear in BRpV, and $a$, $b$ and $c$ are different combinations of the parameters of the model.

It is numerically found that the $b$-term is smaller than the other two, leaving us with two pieces to fit neutrino data. Then, two important conclusions can be learnt from equation \eqref{meff}:

\begin{itemize}

\item The matrix $\boldsymbol{m_{\nu\nu}^{\rm eff}}$ has two non-zero eigenvalues eigenvalues at tree-level, as opposed to BRpV, where radiative corrections are needed to generate a second mass scale.

\item There are two possibilities to fit neutrino data:

\begin{itemize}

\item Case (c1): $\vec \Lambda$ generates the atmospheric scale, $\vec \epsilon$ the solar scale

\item Case (c2): $\vec \epsilon$ generates the atmospheric scale, $\vec \Lambda$ the solar scale

\end{itemize}

\end{itemize}

Since there are no a priori reasons for choosing one of these two options, we will consider both in the following. In any case, the \rpv parameters $\epsilon_i$ and $\Lambda_i$ have to be small in order to predict a correct absolut scale of neutrino mass. For typical SUSY masses order ${\cal O}(100\hskip1mm{\rm GeV})$, $|\vec\Lambda|/\mu^2 \sim 10^{-6}$-$10^{-5}$. If some of the singlet fields are light, i.e.~have masses in the range of ${\cal O}(0.1-{\rm few})$ TeV, also $|\epsilon_i/\mu|$ can be as small as $|\vec\epsilon|/\mu \sim 10^{-6}$-$10^{-5}$.

\section{Neutralino phenomenology}
\label{sec:pheno}

In this section we discuss the phenomenology of a neutralino LSP in s-\rpv at future colliders, focusing on the most important qualitative features. All numerical results shown below have been obtained using the program package SPheno \cite{Porod:2003um}, extended to include the new singlet superfields of this model. For details about the numerical procedure see \cite{Hirsch:2008ur}.

\subsection{Bino vs Singlino}

In principle, it is impossible to give a definite prediction for the nature of the LSP. This is due to the many unknown parameters in the spectra of the seven heavy neutral fermions. But there are general properties, present in most parts of parameter space:

\begin{itemize}

\item There are four states very close to the MSSM neutralinos. In mSUGRA models the lightest of these four states is the Bino, with a mass given by $M_1$.

\item The states $\nu^c$ and $S$ form a quasi-Dirac pair, the so-called Singlino, ${\cal S}_{1,2} \simeq \frac{1}{\sqrt{2}} (\nu^c \mp S)$.

\item The remaining state is the phino, ${\tilde\Phi}$, also present in the NMSSM.

\end{itemize}

We are going to focus on a neutralino LSP, studying two different cases concerning its nature, (i) A Bino-like $\tilde{\boldsymbol{\chi}}_1^0$, as in a typical mSUGRA scenario, and (ii) a Singlino-like $\tilde{\boldsymbol{\chi}}_1^0$, a novel but natural possibility. The parameter that determines in which case we are is the $\nu^c-S$ element in the neutral fermion mass matrix. If it is smaller than the Bino mass the Singlino will be lighter than the rest of neutralinos. Taking the expressions for the mass matrices (see reference \cite{Hirsch:2008ur} for the formulas) one finds that this condition reads $\mathbf{M_{\nu^c S}} = \frac{1}{\sqrt{2}} h v_{\Phi} \lesssim M_1$, what can be naturally achieved.

\subsection{Production}

Since neutrino physics requires that the \rpv parameters are small, the production cross sections are very similar to the corresponding MSSM values \cite{AguilarSaavedra:2005pw}. Therefore, we expect standard decay chains, with gluinos and squarks directly produced at the LHC and a neutralino LSP appearing as part of the \emph{final state} (actually, it will further decay thanks to the breaking of R-parity). For example, a decay chain like

\begin{equation}
{\tilde q} \to q + {\tilde B} \to q + {\cal S}_{1,2} + J \to \dots
\end{equation}

will produce singlinos as an intermediate state, before starting with the \rpv decays.

In conclusion, the production of the lightest neutralino is guaranteed, both for a Bino-like LSP and for a Singlino-like LSP. Moreover, we see that singlinos can be produced at accelerators, although their direct production cross section is negligible, allowing us to study their decays.

\subsection{Decays}

\begin{figure}[htbp]
\begin{center}
\includegraphics[width=0.5\textwidth]{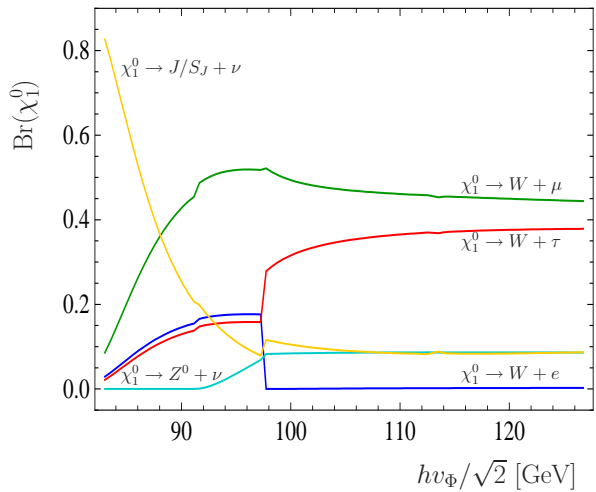}
\end{center}
\caption{Branching ratios for the most important decay modes of the lightest neutralino versus $\frac{1}{\sqrt{2}}hv_{\Phi}$ for a specific, but typical example point, in which the MSSM parameters have been adjusted such that the sparticle spectrum of the standard point SPS1a' is approximately reproduced. The singlet parameters have been chosen randomly, $v_R=v_S=1$ TeV, ${\vec\epsilon}$ and $\vec\Lambda$ have been fitted to neutrino data, such that $\vec\Lambda$ generates the atmospheric scale and ${\vec\epsilon}$ the solar scale.}
\label{fig:decays1}
\end{figure}

With broken R-parity the lightest supersymmetric particle decays. In figure \ref{fig:decays1} we show the most important decay channels for a neutralino with $m_{\chi^0_1} \ge m_{W^{\pm}}$. Note the transition between Singlino LSP (left) and Bino LSP (right) for $\frac{1}{\sqrt{2}}hv_{\Phi} \simeq M_1$ ($M_1 \simeq 98$ GeV in SPS1a' standard point) and how the different branching ratios are modified depending of the nature of the lightest neutralino.

For low values of the LSP mass the channel $J/S_J + \nu$ is usually the most important. The state $S_J$ is a rather light singlet scalar, called the \textit{scalar partner} of the Majoron in \cite{Hirsch:2005wd}, that decays to a pair of Majorons with a branching ratio close to 100\%. Therefore, this channel represents invisible final states, since the Majoron escapes detection, and it has a sizeable branching ratio even for a relatively high $v_R$. We will come back to this important point later. Next in importance are the final states involving $W^{\pm}$ and charged leptons. The relative size of the branching ratios for the final states $W e$, $W \mu$ and $W \tau$ depends on both, (a) the nature of the LSP and (b) the fit to the neutrino data. Finally, the model predicts

\begin{equation}\label{ZtoW}
\frac{\sum_i Br(\chi_1^0\to Z^0 + \nu_i)}{2 \sum_i Br(\chi_1^0\to W^{+} + l_i^{-})} \simeq \frac{g}{4 \cos^2\theta_W}
\end{equation}

with $g$ being a phase space correction factor, with $g \to 1$ in the limit $m_{\chi^0_1}\to\infty$ \cite{Hirsch:2005ag}.

For the case of $m_{\chi^0_1}\le m_{W^\pm}$, one finds that invisible decays also play a very important role, with large branching ratios, which can be even dominant in some parts of parameter space. The rest of decay channels are three body decays mediated by virtual gauge bosons. See \cite{Hirsch:2008ur} for the details.


The induced invisible decays are the most important consequence of the existence of the Majoron. In the previous examples there are regions of parameter space where they are dominant, but still allowing for a sizeable branching ratio to visible final states.

\begin{figure}[htbp]
\begin{center}
\includegraphics[width=0.45\textwidth]{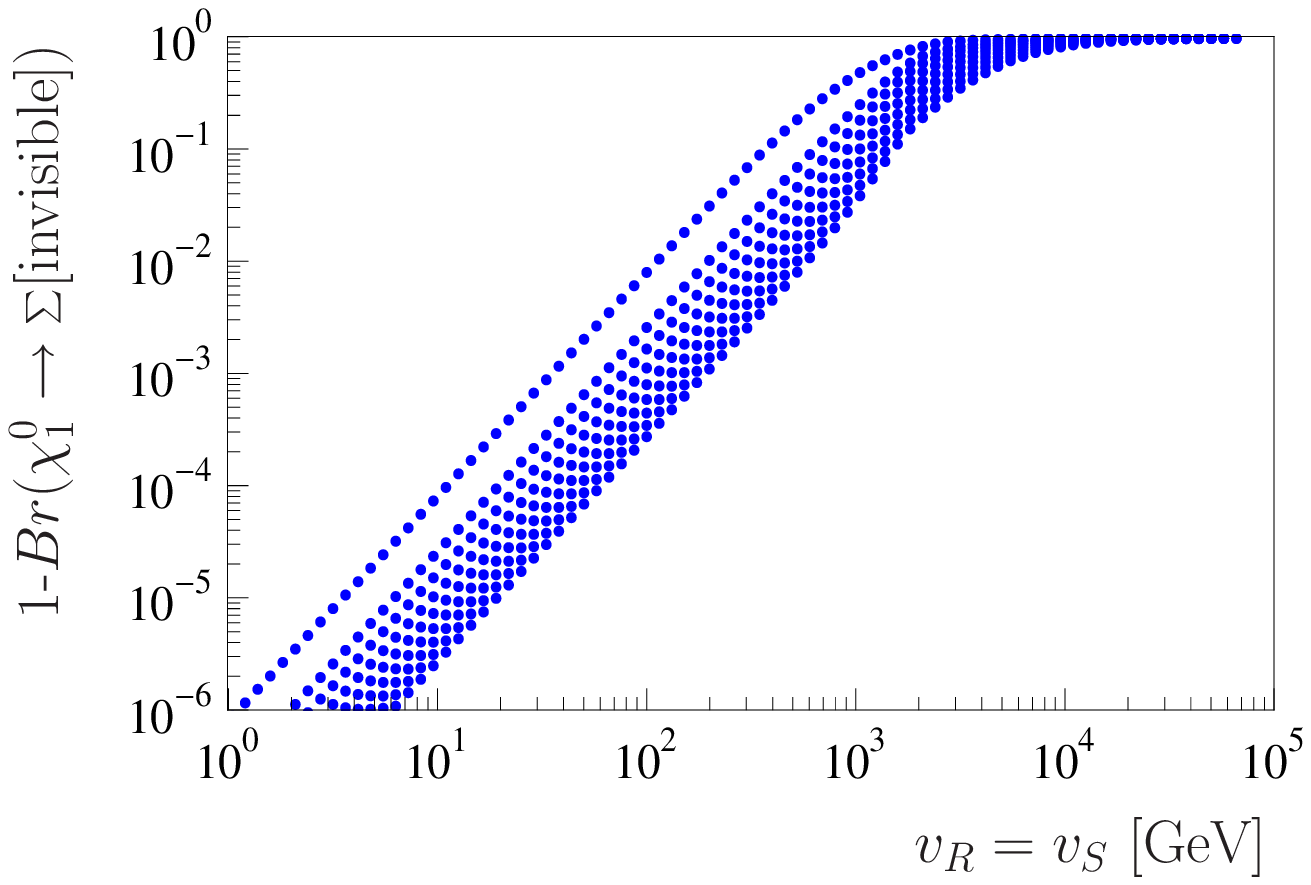}
\includegraphics[width=0.45\textwidth]{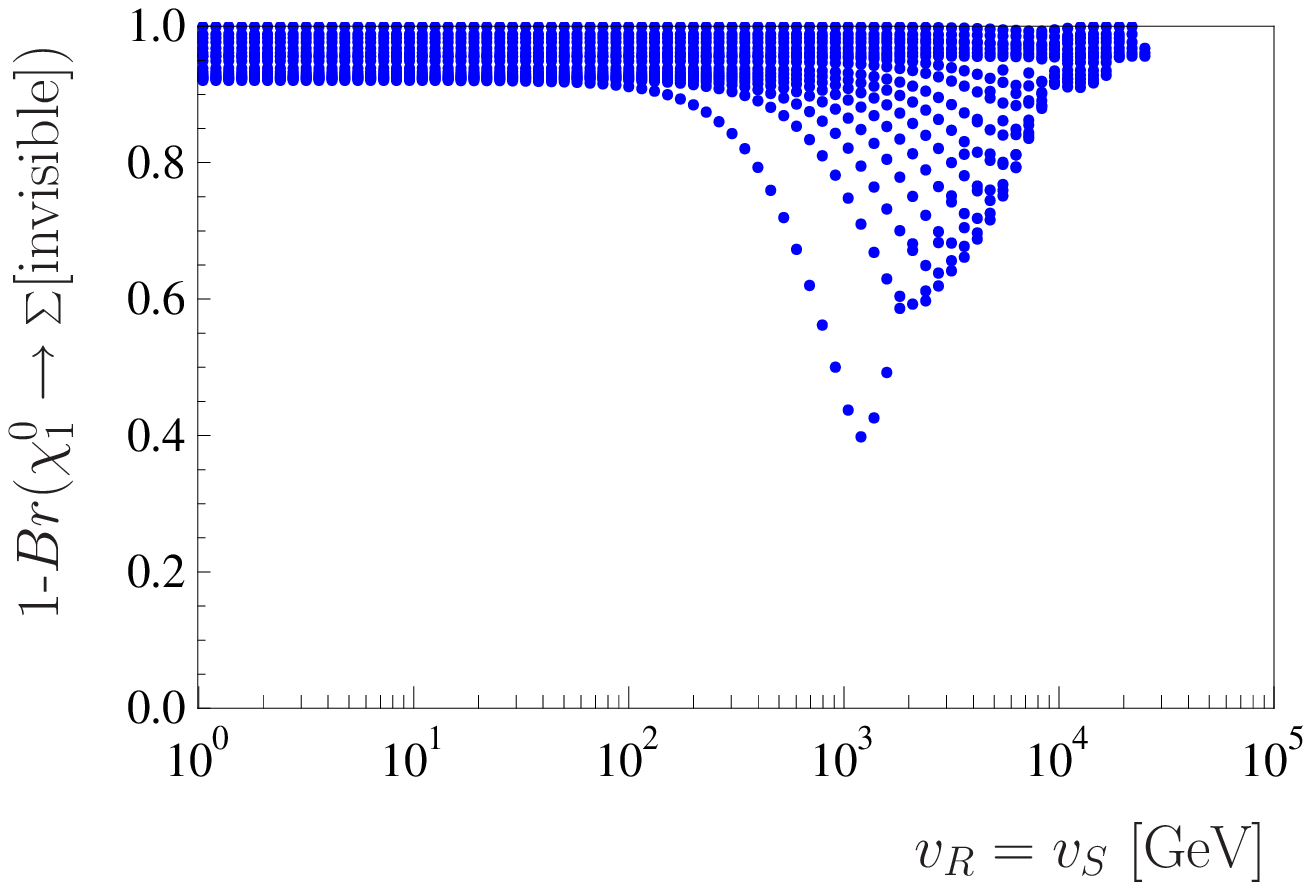}
\end{center}
\caption{Sum over all at least partially visible decay modes of the lightest neutralino versus $v_R$ in GeV, for a set of
$v_{\Phi}$ values $v_{\Phi} = 10-40$ TeV for the mSUGRA parameter point $m_0=280$ GeV, $m_{1/2}=250$ GeV, $\tan\beta=10$, $A_0=-500$ GeV and $sgn(\mu)=+$. To the left $\chi_1^0 \simeq {\tilde B}$; to the right $\chi_1^0 \simeq {\cal S}$.}
\label{fig:Inv_vr}
\end{figure}

Figure \ref{fig:Inv_vr} shows the sum over all at least partially visible decay modes of the lightest neutralino versus $v_R$ in GeV, for a set of $v_{\Phi}$ values $v_{\Phi} = 10-40$ TeV for the mSUGRA parameter point ($m_0=280$~GeV, $m_{1/2}=250$~GeV, $\tan\beta=10$, $A_0=-500$~GeV and sgn$(\mu)=+$). This point was constructed to produce formally a $\Omega_{\chi^0_1}h^2 \simeq 1$ in case of conserved R-parity, much larger than the observed relic DM density \cite{Spergel:2006hy}. The left plot shows the case $\chi_1^0 \simeq {\tilde B}$, the right plot $\chi_1^0 \simeq {\cal S}$. The plot demonstrates that the branching ratio into ${\tilde B} \to J  \nu$ does depend strongly on the value of $v_R$ and to a minor extent on $v_{\Phi}$. Lowering $v_R$ one can get branching ratios for the invisible decay of the ${\tilde B}$ very close to 100\%, thus a very MSSM-like phenomenology, since the Majorons mimic the usual missing $E_T$ signal of the MSSM. This feature is independent of the mSugra parameters, see the correspoding figure in \cite{Hirsch:2006di}. In this case large statistics becomes necessary to find the rare visible neutralino decays, which prove that R-parity is broken. The inconsistency between the calculated $\Omega_{\chi^0_1}h^2$ and the measured $\Omega_{CDM}h^2$ might give a first indication for a non-standard SUSY model. Also, as we will comment in section \ref{sec:exotic}, exotic muon decays are enhanced for low values of $v_R$, and therefore they can provide additional tools to test the model in that region.

Figure \ref{fig:Inv_vr} to the right shows that the case $\chi_1^0 \simeq {\cal S}$ has a very different dependence on $v_R$. We have checked that this feature is independent of the mSugra point. For other choices of mSugra parameters larger branching ratios for Br(${\cal S} \to J + \nu$) can be obtained, but contrary to the Bino LSP case, the sum over the invisible decay branching ratios never approaches 100\%.


Another important property of the decays of the lightest neutralino is their strong correlation with neutrino physics. Since LSP decay and neutrino masses are consequences of the same physical input, the breaking of R-parity, some connection between them is expected. In particular, one can use the decays to $W l_i$ (with the $W$ boson real or virtual) to measure neutrino mixing angles.

\begin{figure}[htbp]
\begin{center}
\includegraphics[width=0.45\textwidth]{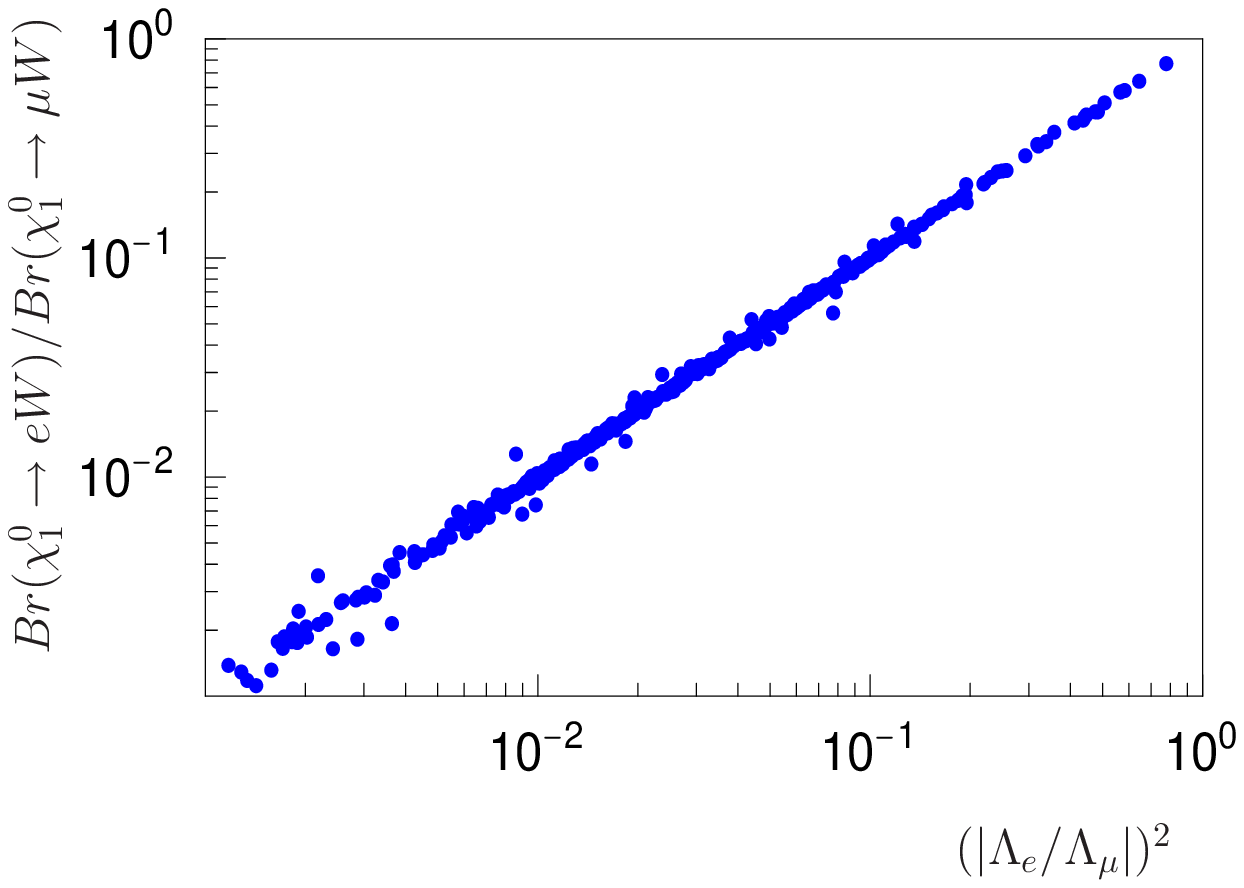}
\includegraphics[width=0.45\textwidth]{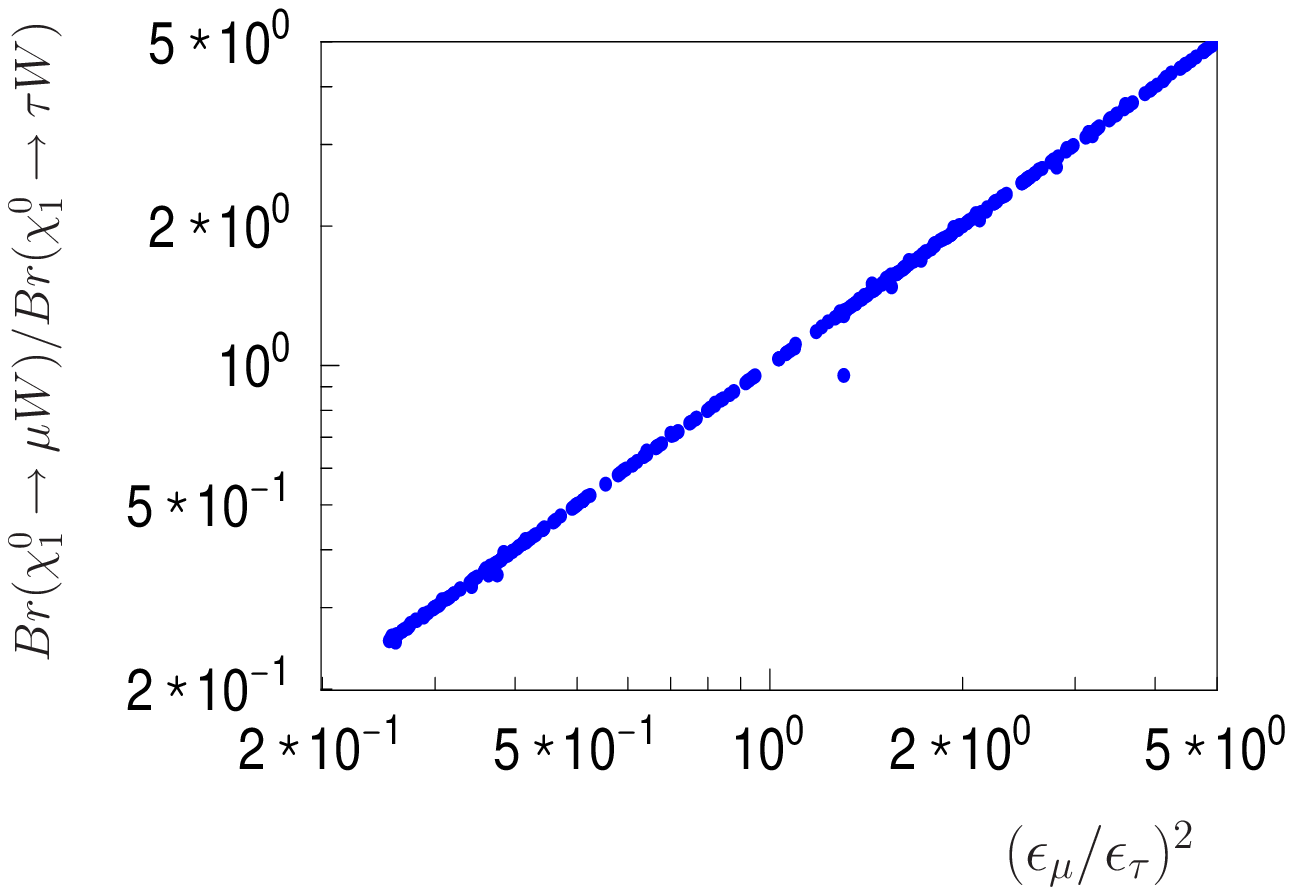}
\end{center}
\caption{Ratios of branching ratios $\tilde{\chi}_1^0 \to W l_i$ and their correlations with \rpv parameters. To the left: Ratio $\frac{Br(\chi^0_1\rightarrow e W)}{Br(\chi^0_1\rightarrow \mu W)}$ versus $(\Lambda_{e}/\Lambda_{\mu})^2$
for a Bino LSP. To the right: Ratio $\frac{Br(\chi^0_1\rightarrow \mu W)}{Br(\chi^0_1\rightarrow \tau W)}$ versus
$(\epsilon_{\mu}/\epsilon_{\tau})^2$ for a Singlino LSP. All points with $m_{LSP} > m_{W}$.}
\label{fig:propor}
\end{figure}

This is possible due to the good approximations that can be made in the $\tilde{\chi}_1^0-W^{\pm}-l^{\mp}_i$ couplings, which are found from the general expressions for the $\tilde{\chi}^0-W^{\pm}-{\tilde \chi}^{\mp}$ vertices (see \cite{Hirsch:2008ur} for explicit formulas). Using these approximations one can show that

\begin{eqnarray}
Br({\tilde B} \to W l_i) &\propto& {\Lambda_i}^2 \\
Br({\cal S} \to W l_i) &\propto& {\epsilon_i}^2
\end{eqnarray}

where the proportionality constant is independent of the generation index 'i' for both cases. This feature is shown in figure \ref{fig:propor}

Since the structure of $\boldsymbol{m_{\nu\nu}^{\rm eff}}$ is given by $\vec \Lambda$ and $\vec \epsilon$, this implies correlations between some combinations of branching ratios and neutrino mixing angles.

\begin{figure}[htbp]
\begin{center}
\includegraphics[width=0.5\textwidth]{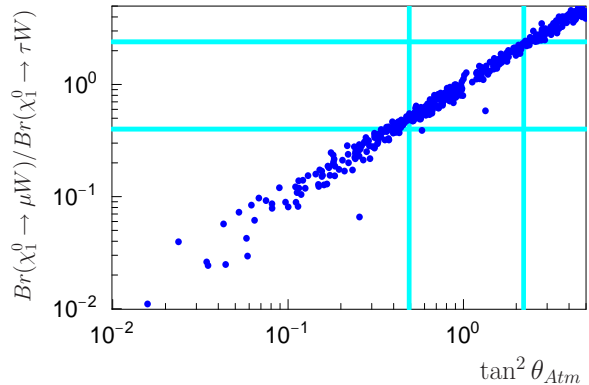}
\end{center}
\caption{Ratio $\frac{Br(\chi^0_1\rightarrow \mu W)}{Br(\chi^0_1\rightarrow \tau W)}$ versus $\tan^2\theta_{Atm}$ for a Bino LSP. Vertical lines are the $3 \sigma$ c.l. allowed experimental ranges, horizontal lines the resulting predictions for the fit (c1).}
\label{fig:plotcorr}
\end{figure}

The correlations depend on the nature of the LSP, Bino or Singlino, and the fit to neutrino data, (c1) or (c2). In figure \ref{fig:plotcorr} we give a representative example, showing the ratio $\frac{Br(\chi^0_1\rightarrow \mu W)}{Br(\chi^0_1\rightarrow \tau W)}$ as a function of $\tan^2\theta_{Atm}$ for a Bino LSP. Using the current experimental range for the atmospheric angle one can predict an allowed range for the ratio, which has to be within $[0.4, 2.1]$ to be fully consistent with the model.

For other cases see \cite{Hirsch:2008ur}. It must be stressed that for the case $m_{\chi^0_1}\le m_{W^\pm}$ the correlations are still present, with virtual $W$ bosons instead of real ones, and similar results can be found with the channels $\tilde{\chi}^0_1 \to l_i qq'$.

In conclusion, these correlations make the model testable at the inminent LHC. However, since we don't know whether case (c1) or case (c2) is realized, the decay of the lightest neutralino is not sufficient to determine the nature of the LSP. We need to reconstruct the complete decay chains and use kinematical variables to obtain some information about the intermediate states.

\section{Exotic muon decays}
\label{sec:exotic}

As it has been shown in section \ref{sec:pheno}, low values of $v_R$ enhance invisible Bino decays, and thus makes it difficult to distinguish the model from conserved R-parity. If Nature has chosen that region of parameter space we will need really high statistics to prove R-parity breaking at colliders. For that reason, it is interesting to have other observables to test the possible existence of the Majoron.

Exotic muon decays involving Majorons \cite{Romao:1991tp}, like $\mu \to e J$ and $\mu \to e J \gamma$, might provide such experimental input, since they are also enhanced for low values of $v_R$. This relation can be directly seen from the coupling $l_i-J-l_j$, which has the form

\begin{eqnarray}\label{coupllJ}
O_{RijJ}^{ccp} &=& - \frac{i (h_E)_{jj}}{\sqrt{2} V} \big[ \frac{v_d v_L^2}{v^2} \delta_{ij} + \frac{1}{\mu^2}(C_1 \Lambda_i \Lambda_j + C_2 \epsilon_i \epsilon_j + C_3 \Lambda_i \epsilon_j + C_4 \epsilon_i \Lambda_j) \big]\nonumber \\
O_{LijJ}^{ccp} &=& \big(O_{RjiJ}^{ccp}\big)^*
\end{eqnarray}

and the $C$ coefficients are different combinations of MSSM parameters. Note that the coupling is divided by $V \sim v_R$, and thus it is larger for low values of $v_R$.

By measuring deviations from the standard decays of the muon one can discover these decay channels or, at least, put bounds on the parameters of the model. In particular, if these processes are not observed at low energy experiments, their results will be translated into lower bounds for $v_R$. If those bounds are in conflict with the non-observation of visible decays of the lightest neutralino the model will be in trouble.

Actually, there are current bounds on the branching ratios of the decays $\mu \to e J$ and $\mu \to e J \gamma$. On the one hand we have results by experiments studying muon decay and looking for $\mu \to e \gamma$, which indirectly can provide information about these signals. As it was already studied in \cite{Jodidio:1986mz}, a visible peak over the background signal $\mu \to e \nu \bar \nu$ would mean the discovery of $\mu \to e J$. However, the current bound $Br(\mu \to e J) < 2,6 \cdot 10^{-6}$ , taken from that reference, does not directly apply to our model, since the authors consider an isotropic familon emission, which is not the case for a Majoron with parity violating couplings. On the other hand, astrophysics puts a worse but more reliable bound thanks to the possible influence of the Majoron in the cooling of red giants. In \cite{Raffelt:1996wa} it is estimated that the coupling e - e - J has to obey $g_{eeJ} < 10^{-13}$ in order to leave important stellar properties unchanged. This can be translated into an approximate bound $Br(\mu \to e J) < (few) \cdot 10^{-5}$. Finally, for the case of $\mu \to e J \gamma$, the reference \cite{Goldman:1987hy} gives the bound $Br(\mu \to e J \gamma) < 1.3 \cdot 10^{-9}$, although is is restricted to a small part of the phase space.

\begin{figure}[htbp]
\begin{center}
\includegraphics[width=0.5\textwidth]{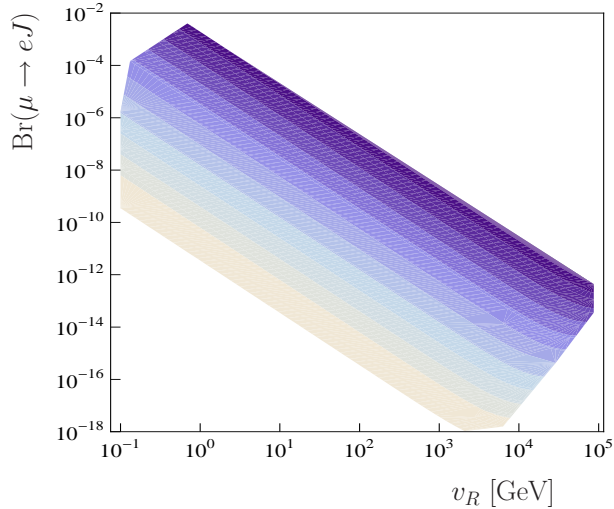}
\end{center}
\caption{$Br(\mu \to e J)$ as a function of $v_R$ for different values of $v_\Phi$. The parameter $v_\Phi$ is taken in the range $[1,75]$ TeV, with darker colours indicating a larger value. The MSSM parameters have been adjusted in the same way as in figure \ref{fig:Inv_vr}. The points have been selected so that neutrino data can be correctly fitted at tree-level.}
\label{fig:vRBr}
\end{figure}

In figure \ref{fig:vRBr} the branching ratio of $\mu \to e J$ is shown as a function of $v_R$ for different values of $v_\Phi$. As previously explained, for low values of $v_R$ the branching ratio can be extremely large, with values clearly above the current bounds. In addition, there is an indirect dependence on $v_\Phi$ through neutrino masses. The parameters $a$, $b$ and $c$ in equation \eqref{meff} are dependent on $v_\Phi$ and they are strongly decreased for high values of this parameter. Then, if $v_\Phi$ is large one needs also high values for the \rpv parameters $\epsilon_i$ and $\Lambda_i$ in order to correctly fit neutrino masses, thus indirectly increasing the coupling \eqref{coupllJ}.

The decay $\mu \to e J \gamma$ can be also of interest for our purposes. Many of the current experiments measuring muon decay are designed to look for photons in the final state, since $\mu \to e \gamma$ is the main goal. Therefore, they are already prepared to fight the background source given by the radiative decay $\mu \to e \nu \bar \nu \gamma$ by using kinematical cuts that suppress undesired signals, although reducing the allowed phase space for $\mu \to e J \gamma$ as well. Moreover, one has to take into account that some experiments, like MEG \cite{Meg:experiment}, are only interested in muon decays including photons, and therefore they are only sensitive to the process $\mu \to e J \gamma$. For more details, see \cite{Hirsch:2009}.

\section{Summary}
\label{sec:con}

We have studied the phenomenology of a model based on the spontaneous breaking of R-parity. In particular, we have concentrated on the decays of the lightest neutralino, showing how we can extract interesting information from their correlations with neutrino physics. These correlations allow us to test the model at the LHC, confronting the values of special combinations of branching ratios with current experimental ranges for the neutrino mixing angles.

In addition, we have shown the impact of the model on muon decays, which can be significantly modified due to the existence of a massless Majoron. Current and future experimental facilities might be able to find a deviation from the standard expectations.

In conclusion, the model is predictive and can be tested at the LHC.

\section*{Acknowledgments}
This work has been done in collaboration with Martin Hirsch, Jochen Meyer and Werner Porod, and is supported by Spanish grants FPA2008-00319 (MEC) and by Acciones Integradas HA-2007-0090. A.V. thanks the Generalitat Valenciana for financial support.

\end{document}